\documentclass[11pt]{article}
\usepackage{moriond,epsfig}

\def\MyPhoto{\psfig{figure=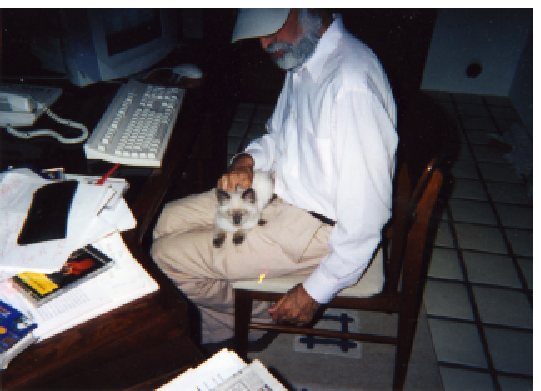,width=60mm}}

\def\NPB{{\em Nucl. Phys.} B}
\def\PLB{{\em Phys. Lett.}  B}
\def\PRL{\em Phys. Rev. Lett.}
\def\PRD{{\em Phys. Rev.} D}

\def\be{\begin{equation}}
\def\ee{\end{equation}}
\def\bea{\begin{eqnarray}}
\def\eea{\end{eqnarray}}

\def\lt{\left}
\def\rt{\right}

\def\gsim{\:\raisebox{-0.5ex}{$\stackrel{\textstyle>}{\sim}$}\:}

\begin{document}
\vspace*{4cm}
\title{THE ``4th GENERATION", $B$-{\it CP} ANOMALIES \& THE
LHC}\footnote{Invited
talk at the EW Moriond '09}.

\author{AMARJIT SONI}

\address{Department of Physics, Brookhaven National Laboratory,\\
 Upton, NY 11973, USA}

\maketitle\abstracts{Attention is drawn to the several 2-3 $\sigma$
``anomalies" in
B, $B_s$ mixings and decays involving CP-observables.  
Perhaps the most interesting theoretical
scenario that could cause such effects is based on warped
extra-dimensional models and indeed some of the effects were predicted
there. However, a rather simple explanation is that based on
a fourth family of quarks with masses in the range of
(400 - 600) GeV.
The
built-in hierarchy of the 4$\times$4 mixing matrix is such that the $t'$
readily provides a needed {\it perturbation} ($\approx 15\%$) to $\sin 2
\beta$ as measured in
$B \to \psi K_s$ and simultaneously is the
dominant source of CP asymmetry in $B_s \to \psi \phi$.
The latter mode is theoretically very clean (unlike the others)
and therefore it would
be extremely desirable that Fermilab gives a very high priority to
clarify this anomaly at the earliest.
4th family explanation allows, with relative
ease, to accomodate the stringent flavor changing constraints which
usually can be quite challenging for new physics scenarios.
Implications for the LHC are briefly discussed; $t'$ and $b'$
lead to high $p_t$ multilepton events, in particular $b'$ decays
produce rather distinctive same sign leptons with asymmetric 
energies.}  

\section{Introduction}


The spectacular performance of the two asymmetric $B$-factories allowed
us to
reach an important milestone in our understanding of {\it CP}-violation
phenomena. For the first time it was established that the observed {\it
CP}-violation in
the $B$ and $K$ systems was indeed accountable by the single, {\it
CP}-odd,
Kobayashi-Maskawa phase in the CKM 
matrix \cite{1963,1973}. In
particular, the
time dependent {\it CP}-asymmetry in the gold-plated $B^0\to\psi K_s$
can be
accounted for by the Standard Model (SM) CKM-paradigm to an accuracy of
around
15\% \cite{HFAG,RMP}. It has then become clear that the effects of a
beyond
the standard model (BSM) phase can only be a perturbation. Nevertheless,
in the
past few years as more data were accumulated and also as the accuracy in
some
theoretical calculations was improved it has become increasingly
apparent that
several of the experimental results are difficult to reconcile within
the SM
with three generations [SM3] \cite{LS07,UT07,LS08}. 
It is clearly
important to follow these indications and to try to identify the
possible origin
of these discrepancies especially since they may provide experimental
signals
for the LHC which is set to start quite soon.
While at this stage many extensions of the SM could be responsible,
it seems that a very simple explanation is provided by the addition of a
fourth family of quarks~\cite{SAARS08,CHOU}.
In fact in~\cite{SAARS08} it is shown that the data suggests that the
charge 2/3 quark of this family needs to have a mass in the range of
(400 - 600) GeV. New physics scenarios with warped extra dimension also
provide a very interesting and viable explanation; indeed some
of the effects were predicted in these models. We will briefly discuss
these as well.

\section{B-CP anomalies}

Below we  briefly summarize  the experimental observations involving
$B$, $B_s$-{\it CP}
asymmetries that are indicative of possible difficulties for the CKM
picture of
{\it CP}-violation.
\begin{enumerate}

\item{The predicted value of $\sin 2\beta$ in the SM seems to be about
2-3
$\sigma$ larger than the directly measured values. Using only
$\epsilon_k$ and
$\Delta M_s/\Delta M_d$ from experiment along with the necessary
hadronic
matrix elements, namely kaon ``$B$-parameter" $B_K$ and using $SU(3)$
breaking
ratio $\xi_s \equiv {f_{bs}\sqrt{B_{bs}}\over f_{bd}\sqrt{B_{bd}}}$,
from the
lattice, alongwith $V_{cb}$ yields a prediction, $\sin
2\beta^{prediction}_{no
V_{ub}} = 0.87 \pm 0.09$ \cite{LS08} in the SM; see Fig.
\ref{fig:SM_fit_novub}. If along with that $V_{ub}\over V_{cb}$ is
also included as an input then one gets a somewhat smaller central value
but
with also appreciably reduced error: $\sin 2\beta^{prediction}_{full
fit} = 
0.75 \pm 0.04$; see Fig.~\ref{fig:SM_fit_withvub}}

\begin{figure}[htb]   
\centering  
\psfig{figure=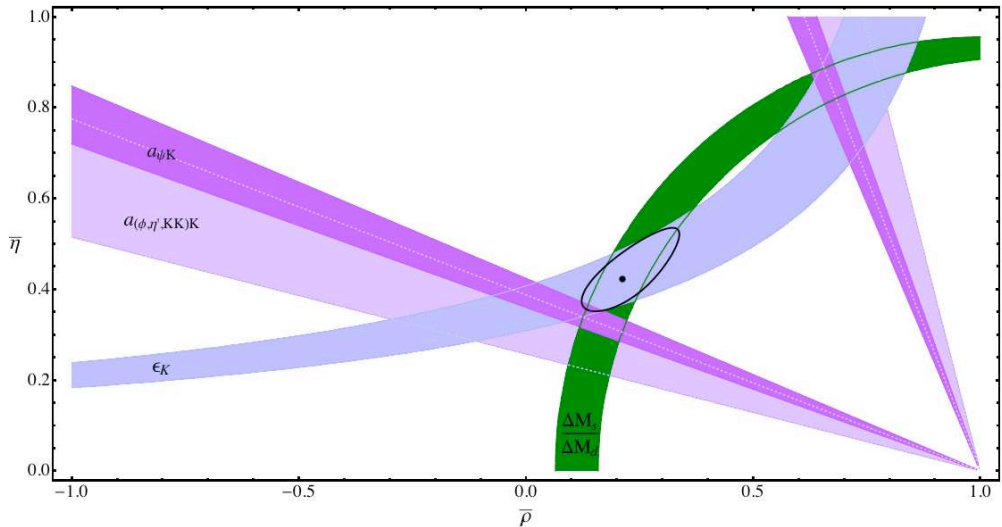,height=3in}
\caption[]{Unitarity triangle fit in the SM. All constraints are 
imposed
at the 68\% C.L.. The solid contour is obtained using the constraints
 from $\varepsilon_K$ and $\Delta M_{B_s}/\Delta M_{B_d}$. The
 regions
  allowed by $a_{\psi K}$ and $a_{(\phi+\eta^\prime + 2 K_s)K_s}$ are
   superimposed; see~\cite{LS08,LS09} for more details.}  
\label{fig:SM_fit_novub}
\end{figure}

\begin{figure}[htb]     
\centering
\psfig{figure=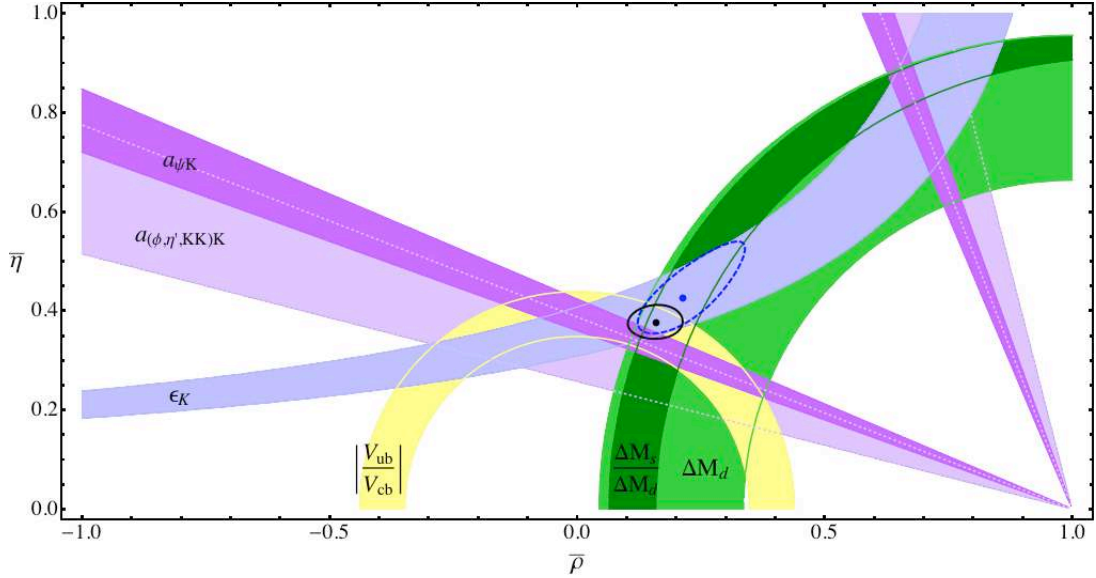,height= 3in}
\caption[]{Unitarity triangle fit in the SM. All constraints are 
imposed at the 68\% C.L.. The solid contour is obtained using the
constraints from $\varepsilon_K$, $\Delta M_{B_s}/\Delta M_{B_d}$
and $|V_{ub}/V_{cb}|$. The dashed contour shows the effect of
excluding $|V_{ub}/V_{cb}|$ from the fit.
The regions allowed by $a_{\psi K}$ and $a_{(\phi+\eta^\prime + 2
K_s)K_s}$ are superimposed;  see~\cite{LS08,LS09} for more
details.}   
\label{fig:SM_fit_withvub}
\end{figure}

\item{The celebrated measurement, via the ``gold-plated" mode $B\to \psi
K_s$,
gives $\sin 2\beta_{\psi K_s} = 0.672 \pm 0.024$ which is smaller than
either
of the above predictions by $\approx$ 1.7 to 2.1 $\sigma$ 
\cite{LS08}.}

\item{As is well known penguin-dominated modes,
such as $B\to (\phi,\eta',\pi^0, \omega, K_sK_s,.
..)K_s$ also allow an experimental determination of $\sin 2\beta$ in the
SM.
\cite{GW97,LS97}.
This method is less clean as it has some hadronic uncertainty.
which was naively
estimated to be at the level of $5\%$ \cite{LS97, GWI97}.
Unfortunately, this
uncertainty cannot
be reliably determined in a model-independent manner. However, several
different estimates \cite{LS24-27} find that amongst these modes,
($\phi$, $\eta'$, $K_s$$K_s$)$K_s$ are rather clean up to an error of
only a
few percent. In passing, we note also another
intriguing feature of many such penguin-dominated modes is that the
central
value of $\sin 2\beta$ that they give seems to be
below the two SM predicted values given above in \#1 and in fact,
in many cases, even below
the value measured via $B\to\psi K_S$ (given in \#2).

To further exhibit this more clearly we show a direct comparison between
the fitted value of $\sin 2 \beta$ and the one obtained by direct
measurements
via $\psi K_s$ and via the penguin dominated modes; see Fig.~
\ref{fig:s2b_comp}. A crucial test of the CKM paradigm of CP violation is
that the single CP-odd phase in the 3$\times$3
mixing matrix must be able to explain {\it simultaneously} the CP
violation
in K-decays as well as in B, $B_s$ decays. This means that
directly measured values
of $\sin 2 \beta$ whether they are obtained from $\psi K_s$ or
from clean penguin modes such as $\eta' K_s$ or $\phi K_s$ must agree
with the fitted value of $sin 2 \beta$. We see that whether we use
$V_{ub}$
or not the fitted value is not only high compared to the $\psi K_s$ one
but
in fact is higher even more compared to the penguin modes. It is also
striking that the central value of the $\sin 2 \beta$ from almost all
ten
or so penguin modes is lower than the fitted value. The only exception
to this is
the mode $B_d \to \phi \pi^0 K_s$ but here the negative error is huge,
O(50\%), so its meaningless to pay attention
to the central value right now.

We also want to emphasize that, strictly speaking, a comparison
of $\sin 2 \beta$ from $\psi K_s$ with the one from penguin-modes
need not reveal New Physics (NP) beyond CKM if NP is confined to $B_d$
mixing
only and does not affect b $\to$ s penguin transtions or b $\to$ c $\bar
c$ s tree decay. Therefore, direct comparison
with the fitted value of
$\sin 2 \beta$
obtained by using $\epsilon_K$ is
essential}.

\begin{figure}[htb]
\centering
\psfig{figure=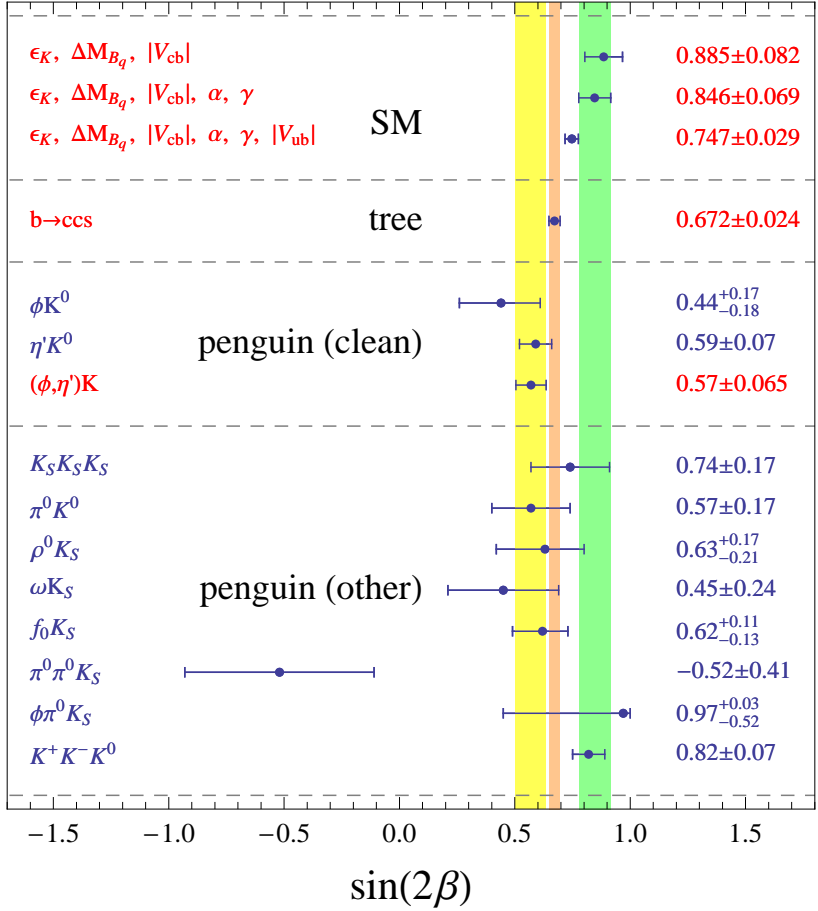,height=5in}
\caption[]{Results of the fit to the unitarity triangle within  
the
SM. In
the table on top we collect the results we obtain for different
selection of inputs. The plot is a graphical comparison between the SM
predictions given above and the direct determinations in $b\to c\bar c
s$ and $b\to s$ penguin modes. In bottom-right table we show the
deviation of the clean $\sin 2\beta$ measurements from the SM
predictions obtained using $\varepsilon_K$, $\Delta M_{B_q}$, $|V_{cb}$,
$\alpha$, $\gamma$. The last column shows the impact of $|V_{ub}|$;
taken from~\cite{LS09}.}  
\label{fig:s2b_comp}
\end{figure}

\item{ Another apparent difficulty for the SM is understanding the
rather large
difference in the direct {\it CP} asymmetries $\Delta A_{CP} \equiv 
A_{CP}(B^-\to K^-\pi^0) - A_{CP}(\bar{B^0}\to K^-\pi^+)= (14.4\pm
2.9)\%$
\cite{HFAG}. Naively this difference is supposed to be zero,
since the two decaying B's are related by switching of the spectator u
and d quark, but it is not by
almost five $\sigma$. Using QCD
factorization \cite{BBNS} in conjunction with any of the four
scenarios for $1/m_b$ corrections that have been 
proposed \cite{BN03},
we were
able to estimate $\Delta A_{CP} = (2.2 \pm 2.4)\%$ 
\cite{LS09} which is
several $\sigma$'s away from the experimental observations. It is
important to
understand that by varying over those four scenarios one is actually
spanning
the space of a large class of final state interactions; therefore the
discrepancy with experiment is serious \cite{FJPZ}. Note also that this
result
for the $\Delta A_{CP}$ is quite consistent with results obtained
when a completely different approach is used to estimate final
state rescattering effects giving rise to the non-leading
corrections~\cite{ccs1}.
However, given our limited
understanding of hadronic decays still makes it difficult to draw
compelling
conclusions from this difficulty for the SM (with three generations).}

\item{ Finally, recently the possibility of the need for a largish
non-standard {\it CP}-phase has been raised 
\cite{LN07,UT08} in the
study of
$B_s\to \psi\phi$
at Fermilab by CDF \cite{cdf} and D0 
\cite{D0} experiments. Since the
presence of a beyond the SM {\it CP} odd phase in $b\to s$ transitions
as
(for example) already emphasized in \cite{LS07}, such non-standard
effects in
$B_s$ decays are quite unavoidable.

Given that this effect ({\it i.e} a non-vanishing CP-asymmetry in $B_s
\to \psi \phi$) is theroretically very clean, it has special importance.
Unfortunately,
at present it is quite limited by statistics and stands at around
$\approx$
2.2 $\sigma$. Fortunately the Tevatron is running very well and
accumulating
luminosity. If analysis with increased luminosity lead to an enhancement
of the significance of this asymmetry, that could be very important.
Thus running
the Tevatron to resolve this issue deserves a very high priority.}

\end{enumerate}

\section{Theoretical interpretations of the anomalies}

In the following we discuss a few theoretical scenarios that could be
the underlying cause for the observed anomalies. However, we first very
briefly consider a model independent point of view and then some
specific models that
appear to be relevant.

\subsection{Brief Summary of the model independent analysis}

One of the important issue is how these B-CP anomalies will impact
search for New Physics at the LHC. For this purpose it is vitally
important to
understand
the scale of NP. With this in mind we~\cite{LS09}
write down dimension-6 operators under the general assumptions of NP in
$\Delta Flavor=2$ effective Hamiltonian for K, $B_d$
or $B_s$ mixing or for the case of $\Delta Flavor=1$
Hamiltonian relevant for
$b \to s$ penguin transitions~\cite{LS09}.  
Table 1  
summarises results of such a model independent analysis. 
We see that the 
scale of
NP is only a few hundred GeV if it orginates from $b \to s$, $\Delta
Flavor=1$ penguin Hamiltonian. It rises to about a few TeV if it
originates from $B_d$ and/or $B_s$ mixing. From the perspective of LHC
the scneario that is the most pessimistic, with NP scale  in the range
of a few tens of TeVs, is when all of the NP resides only in the
dimension-6
LR-operator relevant for the $K -\bar K$
mixing~\cite{BBS_LR}. But in this case the indications for a
non-vanashing  CP-asymmetry in
the CDF and D0 data for $B_s \to \psi \phi$ or for the large value
of $\Delta A_{CP}[K \pi]$ or for the smallish value of $\sin 2 \beta$
from
penguin modes becomes difficult to reconcile.

\begin{table}[t]
\caption{Scale of New Physics, accompanied by a CP-odd phase,
relevant to different scenarios to account for CP anomalies
in the data; taken from Ref. 16.}      
\label{scale}
\begin{center}
\begin{tabular}{|c|c|c|c|} \hline
Scenario  & Operator & $\Lambda\; ({\rm TeV})$ & $\varphi \;({}^{\rm o})$ \\ 
\hline
\vphantom{$ \cases{a \\ b\\ x} $ }
$B_d$ mixing & $O_1^{(d)}$ &
$\cases{1.1 \div 2.1 & no $V_{ub}$ \\ 1.4 \div 2.3 & with  $V_{ub}$}$ &
$\cases{ 15 \div 92 & no $V_{ub}$ \\ 6 \div 60 & with  $V_{ub}$}$ \\
\hline
\vphantom{$ \cases{a \\ b\\ x} $ }
$B_d = B_s$ mixing & $O_1^{(d)} \; \& \; O_1^{(s)}$ &
$\cases{1.0 \div 1.4 & no $V_{ub}$ \\ 1.1 \div 2.0 & with  $V_{ub}$}$ &
$\cases{ 25 \div 73  & no $V_{ub}$ \\ 9 \div 60 & with  $V_{ub}$}$ \\
\hline
\vphantom{$ \cases{a \\ b\\ x} $ }
$K$ mixing &
$\begin{array}{l} 
O_1^{(K)} \\ 
O_4^{(K)} \\ 
\end{array}$ &
$\begin{array}{l} 
< 1.9 \\  
< 24 \\ 
\end{array}$ &
$130\div 320$ \\ \hline
\vphantom{$ \cases{a \\ b\\ x} $ }
${\cal A}_{b\to s}$ &
$\begin{array}{l} 
O_{4}^{b\to s} \\  
O_{3Q}^{b\to s} \\ 
\end{array}$ &
$\begin{array}{l} 
.25 \div .43 \\  
.09 \div .2  \\ 
\end{array}$ &
$\begin{array}{l} 
0 \div 70 \\  
0 \div 30 \\ 
\end{array}$ \\
\hline
\end{tabular}
\end{center}
\end{table}

\subsection{Warped Flavordynamics \& duality}

Perhaps the most interesting and even compelling BSM scenario is that of
warped extra
dimensional
models\cite{RS99} as they offer a simultaneous resolution to EW-Planck
hierarchy
as well as flavor puzzle~\cite{GN99,GP00}. While explicit flavor models are still
evolving,
potentially this class of models does have extra
CP-violating phase(s)~\cite{APS1}
that can have important repercussions for flavor physics.
Indeed in the simplest scenario (with the assumption of ``anarchic''
Yukawas {\it i. e.} in the 5D theory the entries in the Yukawa matrices
are roughly of similar order) it was {\it predicted}~\cite{APS2} that
there should be
 smallish ({\it i.e.} O(20\%)) deviations from the SM
 in $B_d$ decays to penguin-dominated final states such as $\phi K_s$,
 $\eta' K_s$ etc as well as the possibility of largish CP-odd
 phase in $B_s$ mixing which then of course has manifestations
 in  {\it e.g.} $B_s \to  \psi \phi$; forward-backward asymmetry in $X_s
 l^+ l^-$
 etc\cite{APS1,APS2}. However, in this original study, for simplicity
 it was assumed that
 $B_d$ mixing was essentially described by the SM.
 More recently there have been two extensive studies of the possibilty
 of warped
 models being the origin of the several hints in B,
 $B_s$ decays mentioned above\cite{AJB081,AJB082,MN08}.

 A common feature of  these warped models is that they also imply
 the existence of various Kaluza-Klein states, excitated counterparts of
 the gluon, weak gauge bosons and of  the graviton with masses heavier
 than
 about 3 TeV~\cite{adms03}. Note also that unless  the masses of these
 particles are less than about 3 TeV
 their direction detection at LHC will be very
 difficult~\cite{ABKP06,LR_KKG07,shri1,shri2,LR07, ADPS07,AAS08,AS08,APS06}.

 An extremely interesting subtelty about these 5-dimensional warped
 modeles is that they are supposed to be dual to some 4-dimensional
 models with strong dynamics~\cite{JMM97,NIMA00,Rattazi00,Contino03}.
 This motivates us to search for effective 4-dimensional models that
 seem to provide a good description of the data.

\subsection{Two Higgs doublet model for the top quark}
Another interesting BSM scenario that was studied in the context of such
hints of new physics in $B_d$, $B_s$ physics is that of a two higgs
doublet model. Such class of models are of course a very simple
extension
of the SM and in fact in SUSY they find a natural place as there the
2HDM's become a necessity. However, a specific such model, the two
higgs doublet model for the top-quark (T2HDM)\cite{DK} is also of
interest for
a variety of reasons. It naturally accomodates $m_t >> m_b$ by
postulating
that the second Higgs doublet has a much larger VEV ($v_2$) compared
to the VEV ($v_1$) of the ``first" doublet that couples with all the
fermions
other than the top quark. This of course means that this model does
not preserve ``natural flavor conservation"~\cite{GW77}; so, indeed it
has
tree-level flavor
changing-Higgs couplings. However, these are restricted to the charge +
2/3 sector only. Thus the severe $K- \bar K$, $B_d -\bar B_d$, $B_s
-\bar
B_s$
constraints are  respected but the model predicts enhanced
$D^0-\bar D^0$ mixing\cite{WS}
and enhanced flavor changing  decays of the type
Z$\to$ b $\bar s$, t $\to$  c Z etc\cite{ABES}.
Furthermore, the model has additional CP violating phases that can have
many interesting effects in flavor physics\cite{KSW_rad,LS07}

Specifically, the relevance of this model for some of the aforementioned
anomalies in B decays, was examined in ~\cite{LS07}.

\subsection{Possible relevance of a ``4th generation" of quarks}

Interestingly
SM with a fourth generation [SM4]
provides a rather simple explanation for many of the observed deviations
in B, $B_s$ decays
from the predictions of Standard Model's CKM-paradigm, the SM3 (Standard
Model with 3 generations). Indeed the data is suggestive for
the need of a $t'$ in the range (400 - 600) GeV~\cite{SAARS08}.

For completeness,  let us note that SM4 is a simple extension of SM3
with additional up-type ($t'$)  and
down-type ($b'$) quarks.  It retains all the features of the SM.  The
$t'$  quark, like u, c, t quarks,  contributes in the $b \to s$
transition at
the loop level  \cite{HWS87}. The addition of fourth generation means
that the quark
mixing matrix will become a $4 \times 4$ matrix and the parametrization
of
this unitary matrix requires six real parameters and three phases.
The two extra phases imply the possibility
of extra sources of {\it CP} violation \cite{HSS87}. In order to
constrain
these extra parameters we
use experimental inputs from processes such as, $B_d - \bar{B_d}$ and
 $B_s - \bar{B_s}$ mixing, ${\cal{BR}}(B\to X_s \gamma)$,
 ${\cal{BR}}(B\to X_s 
 \ell^+ \ell^-)$ ,
 indirect {\it CP} violation in $K_L \to \pi\pi$ described
 by $|\epsilon_k|$, Z $\to b \bar b$ etc. Table~\ref{tab:inputs}
 summarizes complete list of inputs
 that we have used to constrain the SM4 parameter space. Using these
 input
 parameters we have
 obtained the constraints on various parameters of
 the 4$\times$4 mixing matrix. In Table~\ref{tab:cons}
 we present the one sigma allowed
 ranges of $|V^{\ast}_{t's}V_{t'b}|$ and $\phi_s'$ (the phase of
 $V_{t's}$), which follow from our
 analysis~\cite{mc_oblique}.

\begin{table}[t]
\caption[]{Inputs used to constrain the SM4 parameter space;
the error on $V_{ub}$ is increased (to 2 $\sigma$) to reflect the
disagreement between the inclusive and exclusive methods; taken
from~\cite{SAARS08}.}
\label{tab:inputs}
\begin{center}
\begin{tabular}{|l|}
\hline
$B_K = 0.72 \pm 0.05$ \\
$f_{bs}\sqrt{B_{bs}} = 0.281 \pm 0.021$  GeV \\
$\Delta{M_s} = (17.77 \pm 0.12) ps^{-1}$\\
$\Delta{M_d} = (0.507 \pm 0.005) ps^{-1}$ \\
$\xi_s = 1.2 \pm 0.06$\\
$\gamma = (75.0 \pm 22.0)^{\circ} $ \\
$|\epsilon_k|\times 10^{3} = 2.32 \pm 0.007$\\
$\sin 2\beta_{\psi K_s} = 0.672 \pm 0.024$\\
${\cal{BR}}(K^+\to \pi^+\nu\nu) = (0.147^{+0.130}_{-0.089})\times
10^{-9}$\\
${\cal{BR}}(B\to X_c \ell \nu) = (10.61 \pm 0.17)\times 10^{-2}$\\
${\cal{BR}}(B\to X_s \gamma) = (3.55 \pm 0.25)\times 10^{-4}$\\
${\cal{BR}}(B\to X_s \ell^+ \ell^-) = (0.44 \pm 0.12)\times 10^{-6}$ \\
( High $q^2$ region )\\
$R_{bb} = 0.216 \pm 0.001$\\
$|V_{ub}| = (37.2 \pm 5.4)\times 10^{-4}$\\
$|V_{cb}| = (40.8 \pm 0.6)\times 10^{-3} $\\
$\eta_c = 1.51\pm 0.24$ \cite{uli1} \\
$\eta_t = 0.5765\pm 0.0065$ \cite{buras1}\\
$\eta_{ct} = 0.47 \pm 0.04$ \cite{uli2}\\
$m_t = 172.5$ GeV\\
\hline
\end{tabular}
\end{center}
\end{table}

The SM3 expressions for $\epsilon_k$ and $Z\to b\bar{b}$ decay width
have been
taken from \cite{buras2} and \cite{zbb} respectively whereas the
relevant
expressions for $\Delta{M_s}\over \Delta{M_d}$, along with the other
observables can be found in \cite{buras3}. The corresponding expressions
in
SM4 i.e., the additional contributions arising due to $t'$ quark can be
obtained by replacing the mass of $t$-quark by $m_{t'}$ in the
respective
Inami-Lim functions.
For concreteness, we use the parametrization
suggested
by Botella and Chau
in \cite{chau} for 4$\times$4 CKM matrix
[$V_{CKM4}$]. In $\Delta{M_d}$ and $\Delta{M_s}$, apart from the other
factors, we have the CKM
elements $V_{tq}V^{\ast}_{tb}$
and  $\Delta{M_s}$ respecitively,
contains the term
$V_{ts}V^{\ast}_{tb}$
(with q=d or s),
\bea
V_{tq}V^{\ast}_{tb} &=& -(V_{uq}V^{\ast}_{ub} + V_{cq}V^{\ast}_{cb}
+V_{t'q}V^{\ast}_{t'b} )
\nonumber \\
V_{ts}V^{\ast}_{tb} &=& -(V_{us}V^{\ast}_{ub} + V_{cs}V^{\ast}_{cb}
+V_{t's}V^{\ast}_{t'b} ),
\label{unitary}
\eea
which is replaced by using the unitarity relation,
$\lambda_u+\lambda_c+\lambda_t+ \lambda_{t'}=0$
where $\lambda_q= V_{qb}V_{qs}^*$.
\noindent using the 4$\times$4 CKM matrix unitarity relations.
The phase of
$V_{td}$ and $V_{ts}$ will also be obtained by using this unitarity
relation.
In this way we have reduced the number of unknown parameters by
using information from known parameters.

With a sequential fourth generation, the effective Hamiltonian
describing, as an example, the decay modes  $B^- \to \pi^0 K^-
$ and $\bar B^0 \to \pi^+ K^-$   becomes
\bea
{\cal H}_{eff} &=&\frac{G_F}{\sqrt{2}}\big[ \lambda_u(C_1^u O_1+C_2^u
O_2  +\sum_{i=3}^{10} C_i^u O_i) \nonumber \\
&{}& + \lambda_c\sum_{i=3}^{10} C_i^c O_i
-\lambda_{t'}\sum_{i=3}^{10} \Delta C_i^{t'} O_i\big] \;,
\eea
where $C_i^q$'s
are the Wilson coefficients, $\Delta C_i^{t'}$'s  are the effective
(t subtracted) $t'$ contributions and $O_i$ are the current-current
operators.
Using the above effective Hamiltonian,
for the $b\to s$ transition and
following \cite{LS07} we use the S4 scenario of QCD factorization
approach  \cite{BN03}
for the evaluation of hadronic matrix elements  and the amplitudes
for the decay modes $B\to \pi K$ and $B\to \phi K_s$ for
$m_{t'}= 400$ , $500$, $600$, $700$ GeV respectively.

Using the ranges of $\lambda^{s}_{t'} \equiv |V^{\ast}_{t's}V_{t'b}|$
and
$\phi_s'$,
the phase of $V_{t's}$
as obtained from the fit for different
$m_{t'}$ (Table~\ref{tab:cons}), the allowed regions in the
$\Delta A_{CP}-\lambda^{s}_{t'}$ plane for different values of $m_{t'}$
were studied.
With the 4th family there is some enhancement
and $\Delta A_{CP}$ up to about 8\% may be feasible which is till
somewhat small compared to the observed value ($14.4 \pm 2.9$)\%. Again,
as we mentioned before  this could be due to the inadequacy of the QCD
factorization model we are using.

\begin{table}[t]   
\caption[]{Allowed ranges for the parameters, $\lambda^s_{t'}$ ($\times
10^{-2}$)
and phase $\phi_s'$ (in degree) for different masses $m_{t'}$ ( GeV),
that has
been obtained from the fitting with the inputs in
Table~\ref{tab:inputs};
taken
from~\cite{SAARS08}.} 
\label{tab:cons}
\begin{center}
\begin{tabular}{|l|c|c|c|l|}
\hline
$m_{t'}$&400 & 500 & 600 & 700 \\
\hline
$\lambda^s_{t'}$ & (0.08 - 1.4)& (0.06 - 0.9)& (0.05 - 0.7) & (0.04 -
0.55) \\
\hline
$\phi_s'$ & -80 $\to$ 80 & - 80 $\to$ 80 & -80 $\to$ 80 & -80 $\to$ 80
\\
\hline
\end{tabular}
\end{center}
\end{table}

\begin{figure}
\centering
\psfig{figure=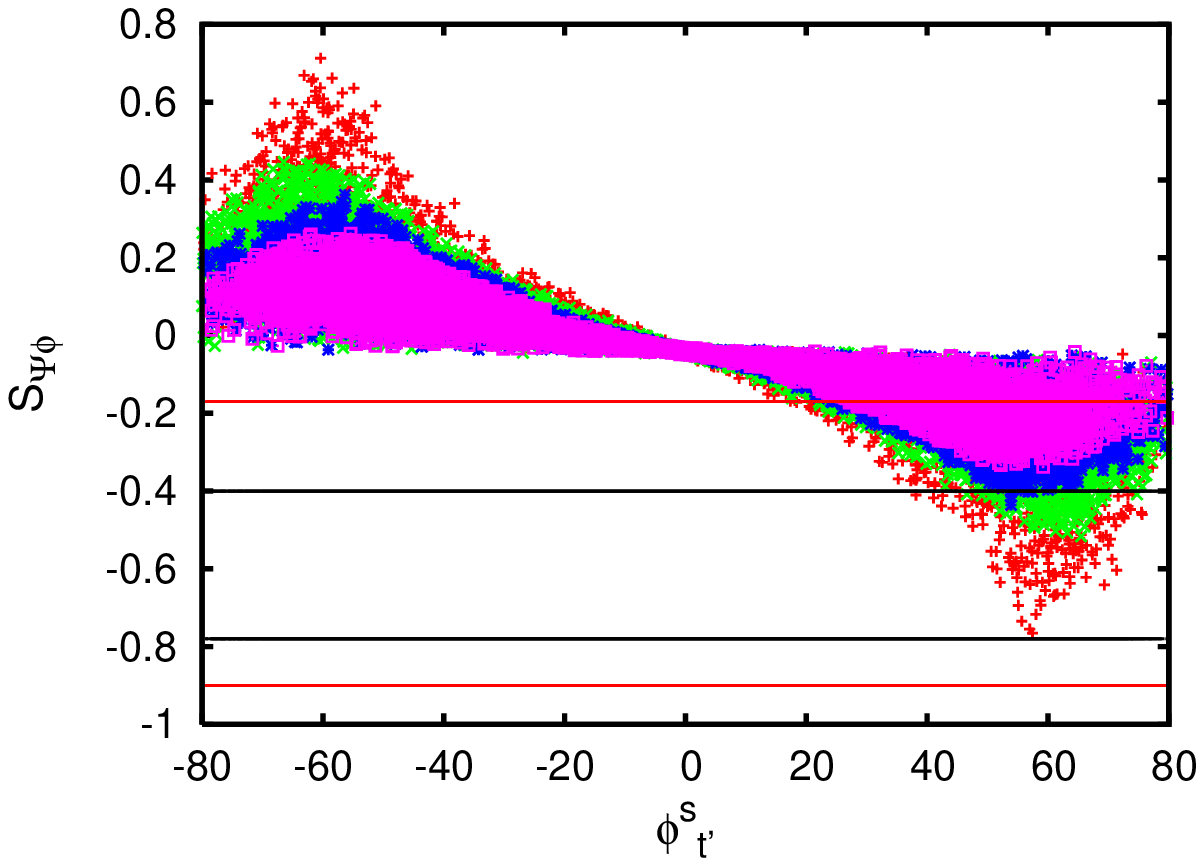,height=1.5in,angle=0}
\psfig{figure=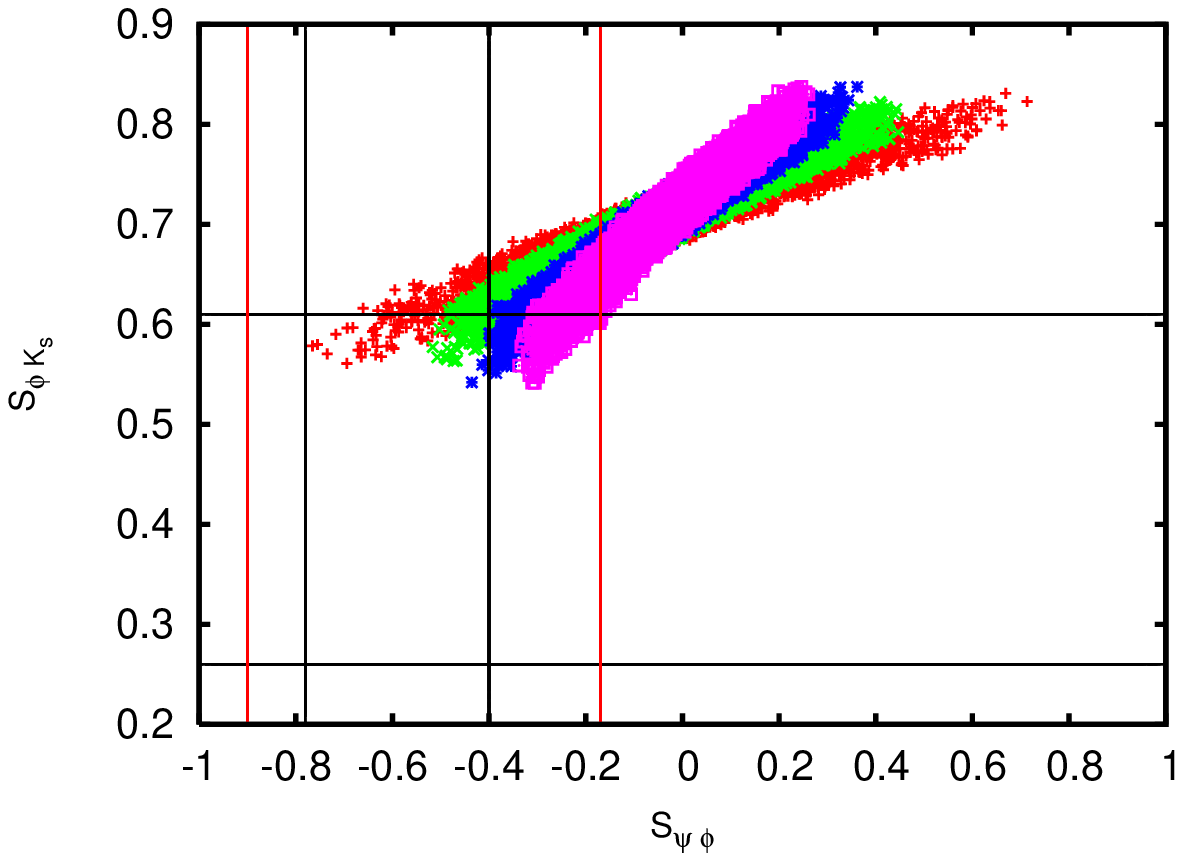,height=1.5in,angle=0}
\caption[]{
The left panel shows the allowed range for $S_{\psi \phi}$ in the
($S_{\psi\phi}-\phi^{s}_{t'}$) plane for $m_{t'}= 400$ (red), $500$
(green),
$600$ (magenta) and $700$ (blue) GeV respectively. Black and red
horizontal
lines  in the figure indicate 1-$\sigma$ and 2-$\sigma$ experimental
ranges
for $S_{\psi\phi}$ respectively.
The right panel shows the correlation between $S_{\phi K_s}$ and
$S_{\psi\phi}$
for $m_{t'}= 400$ (red), $500$ (green), $600$ (magenta) and $700$ (blue)
GeV
respectively. The horizontal lines represent the experimental
$1\sigma$ range for $S_{\phi K_s}$ whereas the vertical lines (Black
1-$\sigma$ and red 2-$\sigma$ ) represent that for $S_{\psi\phi}$; taken
from~\cite{SAARS08}. \label{fig:corr}}.
\end{figure}

In Fig.\ref{fig:corr} we have shown the allowed regions in the
$S_{\psi\phi}-
\phi^{s}_{t'}$ plane for different values of $m_{t'}$ and
also the correlation between
{\it CP} asymmetries in $B\to \psi\phi$ and $B\to \phi K_s$. We follow
the
notation $S_{\psi \phi} = \sin(\phi^{\Delta}_s -2\beta_s) =
\sin2\beta^{eff}_s$
, where $\phi^{\Delta}_s$ is the phase coming from mixing and
$\beta_s = arg(- {V^{\ast}_{tb}V_{ts}\over V^{\ast}_{cb}V_{cs}}) =
1.1^{\circ}
\pm 0.3^{\circ}$, is the phase of $b\to c \bar{c} s$ decay amplitude
\cite{beneke1, uli3}. The range for new $B_s$ mixing phase
$\phi^{\Delta}_s$
is given (@68\% CL) by
$\phi_s^\Delta \,\in \, \lt( -18 \pm 7 \rt)^{\circ}$
or,
$\phi_s^\Delta \, \in \,  \lt( -70 \pm 7 \rt)^{\circ}$.
The corresponding 2-$\sigma$ and 1-$\sigma$ ranges for $S_{\psi\phi}$ is
given
by [-0.90, -0.17] and [-0.78, -0.40] respectively. The large
error on $S_{\phi K_s}$ and $S_{\psi\phi}$ does not allow at present to
draw
strong conclusions on $m_{t'}$, nevertheless the present experimental
bounds
seem to mildly disfavor $m_{t'}>$  600 GeV.

A very appealing feature of the 4th family hypothesis is that it
quite readily
explains the pattern of the observed anomalies.  First of all
the heavy $m_{t'}$ generates a very important new source of electroweak
penguin
(EWP) contribution since, as is well known, these amplitudes are able to
avoid
the decoupling theorem and grow as $m_{t'}^2$ \cite{IL,HWS87}. This
helps to
explain two of the anomalies in $b \to s$ transitions. The enhanced EWP
contribution helps in explaining the difference in CP-asymmetries,
$\Delta A_{CP}$ as it is really the $K^{\pm} \pi^0$ that is enhanced
because of the color
allowed coupling of the Z to the $\pi^0$.
A second important
consequence of $t'$ is that  $b \to s$ penguin has a new CP-odd phase
carried
by $V_{t'b}V^*_{t's}$. This is responsible for the fact that $\sin 2
\beta$
measured in $B \to \psi K_s$ differs with that measured in
penguin-dominated
modes such as $B \to (\phi, \eta', K_s K_s¿)K_s$.

Note also that $\Delta B=2$ box graph gets important new contributions
from the $t'$
since these amplitudes as mentioned before are proportional to
$m_{t'}^2$.  Furthermore,
they are accompanied by new CP-odd phase which is not present in SM3.
This phase is responsible for the fact that the $\sin 2\beta$ measured
in
$B \to\psi K_s$ seems to be  bigger than the value(s) ``predicted" in
SM3 \cite{LS08}
given in item \# 1 on page 1.

Finally, we note briefly in passing how SM4 gives a rather simple
explanation for the size of the new CP-phase effects in $B_d$ versus
$B_s$
mesons. In $B_d$ oscillations resulting in $ B \to \psi K_s$, top quark
plays
the dominant role and we see that the measured value of $\sin2 \beta$
deviates
by $\approx 15\%$ from predictions of SM3.  It is then the usual
hierarchical
structure of the mixing matrix (now in SM4) that guarantees that on
$\sin2 \beta$, $t'$ will only have a subdominant effect.
However,  when we consider $B_s$ oscillations then the role
of $t'$ and $t$ get reversed. In $B_s$ mixing the top quark in SM3 has
negligible CP-odd phase, therein then the $t'$ has a pronounced effect.
SM4
readily explains that just as $t$ is dominant in $\sin 2 \beta$ and
subdominant in $\sin2 \beta_s$, the $t'$ is dominant in $\sin 2 \beta_s$
and
subdominant in $\sin 2 \beta$.

\subsubsection{Improved prospects for baryogenesis}

In SM3 a well known difficulty for baryognesis is that the process is
driven by product of square of mass differences of the 3 up-type quarks
and the 3 down-type quarks:

\bea
Q_B \approx A_{UT-SM3}[{(m_c^2 -
m_u^2)(m_t^2-m_c^2)(m_t^2-m_u^2)(m_s^2-m_d^2)(m_b^2 -
m_s^2)(m_b^2-m_d^2)/m_W^{12}}
\eea

\noindent where $A_{UT-SM3}$ is twice the (invariant) area of the unitarity
triangle for the 3-generation SM\cite{CJ}.
As is well known, $A_{UT-SM3}$
s multiplied by 
an extremely small number making
it very difficult for baryogenesis to occur. However, in SM4 since
each set of 3-generations (of three linearly independent ones)
will have its own associatd CP-odd phase~\cite{1973}.
In the corresponding quantity with new CP-odd phase(s),
therefore,  masses of the first
family no longer need to enter leading to a huge enhancement by the
ratio,

\bea
(m_t^2/m_c^2)(m_t'^4/m_t^4)(m_b^2/m_s^2)(m_b'^4/m_b^4) \approx 10^{16}
\eea

\noindent So, at least from this point of view the chances of baryogenesis in SM4
are
significantly improved~\cite{HOU08,GB97} though its claimed that other
difficulties may still  need to be confronted~\cite{GK08}.

\subsubsection{Repurcussions for SBF}

We now briefly summarize some of the definitive signatures of the 4th
family scenario in flavor observables~\cite{WIP} at the flavor 
facilities of today or tomorrow, such as LHCb and Super-B Factory (SBF).
The need for new CP
phase(s)
beyond the single KM phase\cite{1973} of course must continue to
persist.
This means
that the three values of $\sin 2 \beta$, the fitted one, the one
measured
via $\psi K_s$ and the one measured via penguin dominated modes
({\it e.g.} $\phi K_s$, $\eta'K_s$ etc.) should continue to
differ from each other as more accurate analysis become available.
Furthermore, $B_s$ mixing should also continue to
show the presence of a non-standard phase ({\it e.g.} in $B_s \to \psi
\phi$) as higher statistics are accumulated.
For sure SM4 will have many more interesting
applications in flavor physics, for example, direct CP
asymmetry in $K_L \to \pi^0 \nu \bar \nu$, forward-backward asymmetry
in $B \to X_s l^+ l^-$ etc. which need to be explored~\cite{WIP}.

\subsubsection{Repurcussions for the LHC}

For the LHC,
one definitive pediction of this analysis is a 4th family of quarks
in the range of $\approx$ 400-600 GeV and the
detection of these heavier quarks and their leptonic counterparts
deserves attention.
EW precision constrains the mass-splitting between $t'$ and $b'$ to be
small,
around $50$ GeV \cite{NEED}.
This constraint will clearly have important
repercussions for their decays and therefore their signals at the LHC.

For 500 GeV quarks the cross-section for pair production by gluon fusion
is estimated to be around 4pb~\cite{mgm}. The final states via their
decays should
have pair of opposite sign charged leptons with missing energy carried
by neutrinos
and a $t-\bar t$ pair.
Since the $t^{'}-b^{'}$ mass splitting is likely to be less than $m_W$,
the 2-body
decays of the $t^{'}$ (assuming it is heavier)
may not be kinemattically allowed.
This should have the indirect consequence of boosting BR for the 2-body
mode,
$t{'} \to b W$.

For the $b'$ we should expect prominent 2-body mode, $b' \to t + W$.
The partial width will be suppressed by $V_{b't}$ boosting its lifetime.
Flavor changing $b' \to b Z$ may also be enhanced and quite
interesting.

The decays of $t^{'}$ and/or $b^{'}$ should lead to hgh $P_t$
multilepton
final states. Decays of the $t^{'}$ are likely to be dominated by
$t^{'}$$ \to b + W$, followed by decays of the W. Thus we should expect
opposite sign dilptons from pairs of $t^{'}$. From the pair production
of $b^{'}$ followed by their decays, we can get both opposite sign or
same sign dileptons.
The same sign ones arise ({\it e.g.}) when one lepton comes from $W^-$
originating from
the $b'$ and the other from the $W^+$ originating from the $\bar t$
on the opposite side. These high $p_t$ same sign  dileptons clearly
provide very distinctive signatures. Futhermore, the opposite sign
leptons tend to have similar energies, whereas same sign  tend to be
asymmetric in energy with their average energies differing by a factor
of about two.

Another important consequence of the 4th generation scenario with such
heavy quarks is that correspondingly the higgs may be heavier with
$m_H \gsim 300 GeV$~\cite{kribs_ew}. This could make its detection via
decays to ZZ, WW final
states more prominent and perhaps easier than the light mass case.

\subsubsection{Possibilty for a dark matter candidate}
As far
as the lepton sector is concerned, it is clear that the 4th family
leptons have
 to be quite different from the previous three families in that the
 neutral
 leptons have to be rather massive, with masses $> m_Z/2$. This may also
 be a
 clue that the underlying nature of the 4th family may be quite
 different
 from the previous three families\cite{DM4}. KM \cite{1973} mechanism
 taught us
 the crucial role
 of the three families in endowing CP violation in SM3. It is
 conceivable that
 4th family plays an important role \cite{HOU08,GK08} in yielding enough
 CP to
 generate baryogenesis which is difficult in SM3. Of course it also
 seems highly
  plausible that the heavy masses in the 4th family play a significant
  role in
  dynamical generation of electroweak symmetry breaking. In particular,
  the
  masses around 500 or 600 GeV that are being invoked in our study,
  point to a
  tantalizing possibility of dynamical electroweak symmetry breaking as
  the
  Pagels-Stokar relation in fact requires quarks of masses around 500 or
  600 GeV
  for dynamical mass generation to take place \cite{PS,HHT,BH,BURD,uni}.
  For such heavy masses the values of Yukawa coupling will be large but
  not so
  large to break down perturbation theory.
  Clearly all this brief discussion is signaling is that there is a lot
  of
  physics involving the new family that needs to be explored and
  understood.

  \subsubsection{Does need some tuning}

  While from the point of view of dynamical EW symmetry breaking,
  baryogenesis
  and the possibilty of a dark matter candidate, a pair of
  4th generation-like heavy
  quarks may be quite interesting, to put things in perspective we
  should also
  recall that some degree of cancellations or tuning will be needed to
  accomodate such a family. First of all the EWP constraints show that
  some of the contribution of the heavier quarks to the S and
  T-parameters
  may
  have to cancel against a heavier higgs. Moreover, the EWP constraints
  also
  require that for such heavy quarks, the  mass difference between
  the $t'$ and the $b'$
  cannot be more than around 50 GeV~\cite{kribs_ew}. This means their
  masses may be degenerate
  to about 10\%. While such a tuning may be a cause for concern, it
  may be worth remembering that this degeneracy is a lot less than
  proton-neutron mass difference which we now understand results from
  isospin symmetry of the strong interactions.

\subsubsection{Cannot be a conventional 4th generation}

Since LEP experiments tell us that
there cannot be another squential, essentially masslss neutrino,
the underlying nature of this ``4th generation" that we need for
fixing B-CP anomalies must be quite different. In fact,
from our point of view, all that is really needed is a  heavy
charge + 2/3 quark that paricipates in the EW interactions
through the 4$\times $4 mixing matrix~\cite{hidden}. Thus the underlying nature
of this 4th generation may be quite different from the previous
three\cite{oa09}.

\section{Summary \& Outlook}

Attention is drawn to several 2-3 $\sigma$ indications for new
CP violatng physics in B, Bs and possibly in K-mixings and decays.
Heavier quarks of another (``4th") family offer a simple explanation
for these effects. These indications should be resolved with high
priority. In particular, the hints in $B_s \to \psi \phi$ are very clean
from a theoretical standpoint and since the fermilab Tevatron is also
working very well and acumulating data at a high luminosity, it is
highly desirable to let the Tevatron run for another few years so
that we can get to the bottom of this anomally as soon as possible.

If these signals for NP are confirmed they will obviously have a large
impact on flavor and on collider physics. It is useful to note that
the flavor structure of warped model is sufficiently intricate that
signals such as those highlighted here could originate from
this class of new physics. However, our current understanding
of these models is such that the Kaluza-Klein particles are likely to
have masses
heavier than about 3 GeV making their direct verification at the
LHC rather challenging. In that case, higher sensitivity
experiments in the flavor sector, at high luminosity
machines may be our only way to seek confirmation.
In contrast, if a 4th family explanation is the underlying cause
then the masses of the new particles need only be around 600 GeV making
their detection at LHC also quite feasible.

\section*{Acknowledgements}
I want to thank the organizers of the EW Moriond 2009, and in
particular Jean-Marie Frere for inviting me. Also, I want to thank
my collaborators on the 4th family analysis,
Anjan Giri, Ashutosh Alok, Rukmani Mohanta and
especially Soumitra Nandi and also Enrico Lunghi, Michael Begel,  
Hooman Davoudiasl, Shrihari Gopalakrishna, Thomas Gadfort and Christian
Sturm for 
useful discussions.
This work is supported in part by the US DOE contract No.
DE-AC02-98CH10886.

\end{document}